\documentclass{aa} 
\usepackage{graphics}

\begin{document}

\thesaurus{07(08.01.3; 08.05.1; 08.06.3; 02.12.1; 02.12.3; 02.20.1)}

\title{Fundamental parameters of Galactic luminous OB stars}
\subtitle{V. The effect of microturbulence\thanks
{The INT is operated on the island of 
La Palma by the RGO in the Spanish Obervatorio de El Roque de los Muchachos 
of the Instituto de Astrof\'\i sica de Canarias.}}

\author{M.R. Villamariz\inst{1} \and A. Herrero\inst{1,2}}

\institute{Instituto de Astrof\'\i sica de Canarias, E-38200 La Laguna, 
Tenerife, Spain
\and
Departamento de Astrof\'{\i}sica, Universidad de La Laguna,
Avda. Astrof\'{\i}sico Francisco S\'anchez, s/n, E-38071 La Laguna, Spain
}

\offprints{M.R. Villamariz}
\mail{ccid@ll.iac.es}

\date{15 December 1999 / 14 March 2000}

\titlerunning{The effect of microturbulence}
\authorrunning{M.R. Villamariz \& A. Herrero}

\maketitle

\begin{abstract}

We study the effect of microturbulence in the line formation calculations of
H and He lines, in the parameter range typical for O and early B stars. We
are specially interested in its effect on the determination of stellar
parameters: $T_\mathrm{eff}$, $\log g$ and specially on the He abundance. 

We first analyze the behaviour of H and He model lines between 4\,000 
and 5\,000 \AA~ with microturbulence and find that for O stars
only \ion{He}{i} lines and \ion{He}{ii} $\lambda$ 4686 are sensibly affected
by microturbulence, and that models with lower gravities, the ones suitable
for supergiants, are more sensitive to it. 

Using a test procedure we show that the expected changes in $T_\mathrm{eff}$,
$\log g$ and Helium abundance due to the inclusion of microturbulence in the
 analysis, are small.
We analyze five stars (two late, one intermediate and two early O stars) 
using microturbulence
velocities of 0 and 15 kms$^{-1}$ and confirm the result of the previous test.
The parameters obtained for 15 kms$^{-1}$
differ from the ones at 0 kms$^{-1}$ within the limits of the standard error
box of our analysis. Only later types reduce their He abundance, by 0.02 in
$\epsilon$. Comparing with values
in the literature we find that the range of our changes agree with 
previous results. In some cases other effects can add to microturbulence,
and further reduce the He abundance up to 0.04. 
The quality of the line fits only improves for \ion{He}{i} $\lambda$
4471, but not to the extent of completely solving the so--called dilution
effect. 

Therefore our conclusion is that microturbulence is affecting the derivation
of stellar parameters, but its effect
is comparable to the adopted uncertainties. Thus it can reduce
moderate He overabundances and solve line fit quality differences, but it
cannot explain by itself large He overabundances in O stars.


\end{abstract}

\section{Introduction}

Although microturbulence is a fundamental parameter when deriving abundances
from stellar spectra, it has been systematically ignored when deriving He
abundances from quantitative spectroscopical NLTE analyses of OB stars
(Herrero et al. \cite{Herrero92}, Smith \& Howarth \cite{Smith&Howarth94}),
being the first exceptions to this the works by McErlean et
al. (\cite{McErlean98}) and Smith \& Howarth (\cite{Smith&Howarth98}).

One of the main reasons to ignore it is that the use of NLTE techniques
reduced the need of microturbulence for the reproduction of the observed
metallic line strengths and they even made
abundances less sensitive to the value adopted for microturbulence than in
LTE analyses (Becker \& Butler \cite{Becker&Butler89}).

In quantitative NLTE analyses of OB stars, the lines of H and He are used to
determine stellar parameters (namely, effective temperature,
$T_\mathrm{eff}$, logarithmic surface gravity, $\log g$, and He abundance). Their line
profiles are dominated by the Stark broadening, and microturbulence,
included in the standard way, only adds an extra Doppler width to the thermal
broadening. First values for microturbulence found in LTE for main sequence B
stars, of the order of 5 kms$^{-1}$ (Hardorp \& Scholz \cite{Hardorp&Scholz70}),
showed to be small enough to be of no importance when compared to H and He
thermal velocities, that in such hot atmospheres are well above 5
kms$^{-1}$. This together with the use of NLTE made the influence of
microturbulence in the overall profile seem negligible.

When measuring microturbulence in OB supergiants the situation is quite
different, as its value has always showed to be comparable or well above the
thermal velocities of H and He, and even, in some cases, above the speed of
sound in these atmospheres (Lamers \cite{Lamers72}, Lennon \& Dufton
\cite{Lennon&Dufton86}). This was not changed by later NLTE analyses (Lennon
et al. \cite{Lennon91}, Hubeny et al. \cite{Hubeny91}, Gies \& Lambert
\cite{Gies&Lambert92}, Smartt et al. \cite{Smartt97}), in which the values of
microturbulence obtained are usually reduced but never to values clearly
without contradiction. Kudritzki (\cite{Kudritzki92}) and Lamers \& Achmad
(\cite{Lamers&Achmad94}) explained that this could be due to the presence of
wind outflow in these stars, that can mimic the effect of microturbulence in
the line profiles. 

Most of the previously referred works are on early B and late O
supergiants. Little work has been done on early O stars (see Hubeny et
al. \cite{Hubeny91}), mainly because for them, with scarce metallic lines, it
is very difficult to measure the value of microturbulence. And it is in 
the whole range of O stars that we are interested.

One problem that is systematically found in all analyses of OB spectra is the
imposibility of finding a consistent fit of all \ion{He}{i} lines with a
unique set of parameters. This difficulty appears worse in supergiants than
in main sequence stars, and it is bigger between results from singlet and
from triplet lines, but also appears within different lines in one system. We
call this the {\emph {\ion{He}{i} lines problem}}.

This discrepancy between singlet and triplet lines of \ion{He}{i} was
investigated by Voels et al. (\cite{Voels89}), who considered that atmospheric
extension was responsible and so they called this effect ``generalized
dilution effect''. But recent works with spherical, mass losing models show
that this can not be the only reason (Herrero et
al., \cite{Herrero95}, \cite{Herrero00}).
Furthermore, McErlean et al. (\cite{McErlean98}) and
Smith \& Howarth (\cite{Smith&Howarth98}) find that the inclusion of
microturbulence in the line formation calculations reduces, although not to
all its extent, the discrepancy between the fits of singlet and triplet
lines. They also find better and more consistent fits for the whole set of
\ion{He}{i} lines when microturbulence is considered.

One more reason that encourages us to study the effect of microturbulence is
that a careful inspection of several works shows that even for main sequence
stars,values of microturbulence comparable to the thermal velocities of at
least He are obtained in NLTE (Gies \& Lambert \cite{Gies&Lambert92}, Kilian
et al. \cite{Kilian91}), which can invalidate the hypothesis of 
negligible influence of microturbulence in the profiles.

So in this paper we want to study microturbulence in the range of parameters
typical for O stars of any luminosity class, specifically pointing to its
effect on the determination of stellar parameters and also on the
{\emph {\ion{He}{i} lines problem}}. We are specially interested in investigating
whether the inclusion of microturbulence can solve the {\it He discrepancy}
(Herrero et al. \cite{Herrero92}), as Vrancken (\cite{Vrancken97}), McErlean
et al. (\cite{McErlean98}) and Smith \& Howarth (\cite{Smith&Howarth98})
suggest. This now well known problem has induced a lot of theoretical
improvements in both evolutionary and model atmosphere theories in order to
explain it. Recent works including sphericity and mass loss in the spectral
analysis (Herrero et al. \cite{Herrero95}, \cite{Herrero00}, Israelian et
al. \cite{Israelian99}) show that with these new models, more suitable for
these stars than the plane--parallel hydrostatic ones, the discrepancy is not
solved. Evolutionary models including additional mixing inside the star can
explain enhanced He photospheric abundances during the H-burning phase, this
mixing being induced by different physical mechanisms like
rotation and turbulent diffusion (Dennisenkov, \cite{denn94}, 
Meynet \& Maeder, \cite{Meynet&Maeder97}, Maeder \& Zahn,
\cite{Maeder&Zahn98}), or turbulent diffusion and
semiconvection (Langer \cite{Langer92}). Sometimes   
angular rotational velocity changes during evolution 
are included (Langer \& Heger, \cite{LH98}, Heger, \cite{Heger98}). 
For a recent review, see Maeder \& Meynet (\cite{maemey00}). Therefore
it becomes of great importance to see whether
microturbulence can be responsible for the {\it He discrepancy}.

In Sect. 2 we present line formation calculations of H and He lines and their
behaviour with microturbulence in the O stars domain. In Sect. 3 we determine
the effect of including microturbulence in the determination of stellar
parameters, and in Sect. 4 we analyze some stars with and without considering
microturbulence. Sections 5 and 6 are then dedicated to our discussion and
conclusions, respectively .

\section{Microturbulence in H and He line formation calculations}

In order to study the behaviour of H and He lines with microturbulence, we
perform line formation calculations in the parameter range typical for O to
early B stars of any luminosity class:

	--between 30\,000  and 45\,000 K in $T_\mathrm{eff}$

	--between 3.05 and 4.00 in $\log g$ (in c.g.s. units)

	--for $\epsilon$= 0.10 and 0.20, with $\epsilon=\frac{N(He)}{N(He)+N(H)}$

where N(x) is the number density of atom X.
We follow the classical technique of calculating a NLTE model 
atmosphere of H
and He, in radiative and hydrostatic equilibrium and with plane--parallel
geometry (calculated with
ALI, see Kunze, \cite{Kunze95}), 
and then solve the statistical equilibrium and transfer
equations and perform the formal solution for the lines of H and He
chosen using DETAIL \& SURFACE (Butler \& Giddings
\cite{Butler&Giddings85}). In this final step we have included UV metal
line opacities in the calculations, in order to obtain more realistic
profiles (see Herrero, \cite{Herrero94}, and Herrero et
al., \cite{Herrero00}, for details). As shown in the last reference,
plane--parallel models are unable to reproduce properly spectra of massive OB
stars around 50\,000 K and hotter, that is why we stop our study at 45\,000K. 

Microturbulence is introduced in the standard way, by adding an extra Doppler
width to the thermal broadening of the line, which is then convolved with the
rest of the broadening mechanisms. We have considered it in both the
equations of statistical equilibrium and radiative transfer.
As we don't consider other
motions (turbulent or not) in the determination of the stellar
structure, we prefer to restrict the introduction of microturbulence to the
absorption coefficient. Therefore, we do not include it in the
calculation of the structure of the atmosphere via turbulent
pressure. For similar reasons also we do not even bother about its dependence
 on depth. We don't treat separately the effect on
the populations and line profiles, as this has been recently
studied by McErlean et
al. (\cite{McErlean98}) and Smith \& Howarth (\cite{Smith&Howarth98})
(whose results we support).

We make line formation calculations for the lines we usually consider to
perform our analyses: H$_\gamma$ and H$_\beta$ for \ion{H}{i}; \ion{He}{i}
$\lambda$$\lambda$ 4387, 4922, 4471 \AA~ for \ion{He}{i}; and \ion{He}{ii}
$\lambda$$\lambda$ 4200, 4541, 4686 \AA~ for \ion{He}{ii}.

In order to study the behaviour of these lines with microturbulence, we
perform line formation calculations for microturbulence 
velocity values from 0 to
20 kms$^{-1}$, for different sets of parameters, representing the spectral
types and luminosity classes we are interested in: 

	-- $T_\mathrm{eff}$= 30\,000, $\log g$= 3.05, $\epsilon$= 0.10 for
	late O and early B supergiants. The same
	 with $\log g$= 4.00 for dwarfs. 

	-- $T_\mathrm{eff}$= 35\,000, $\log g$= 3.20, $\epsilon$= 0.10 for
	``middle'' O supergiants. The same with $\log g$= 
	4.00 for dwarfs. 

	-- $T_\mathrm{eff}$= 40\,000, $\log g$= 3.40, $\epsilon$= 0.10 for
	early O supergiants. The same with $\log g$= 4.00 
	for dwarfs. 

	-- $T_\mathrm{eff}$= 45\,000, $\log g$= 3.60, $\epsilon$= 0.10 for
	very early O supergiants. The same with $\log g$= 
	4.00 for dwarfs. 

The values of $\log g$ for supergiants are close to the lowest that could be
converged for every $T_\mathrm{eff}$. For dwarfs of increasing
$T_\mathrm{eff}$, $\log g$ becomes a little below 4.00, but this value still
represents well this luminosity class. Finally, these calculations were
repeated for $\epsilon$= 0.20, to look for differential effects with He abundance.

\begin{figure}

\resizebox{\hsize}{!}{\includegraphics{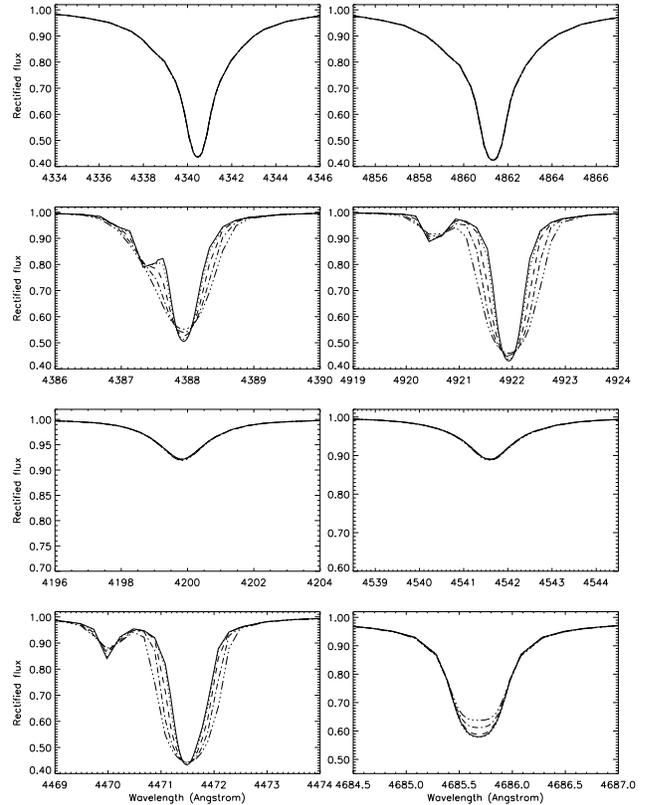}}
\caption[]{From top to bottom and left to right: H$_\gamma$ , H$_\beta$ ,
\ion{He}{i} $\lambda\lambda$ 4387, 4922, \ion{He}{ii} $\lambda\lambda$ 4200,
4541, \ion{He}{i} $\lambda$ 4471, \ion{He}{ii} $\lambda$ 4686, for the model
with $T_\mathrm{eff}$= 30\,000, $\log g$= 3.05, $\epsilon$= 0.10, representing
late O and early B supergiants. Microturbulence takes values of 0 (solid
line), 5 (dotted), 10 (dashed), 15 (dash dotted) and 20 (dash tri-dotted)
kms$^{-1}$.}
\label{fig1}

\end{figure}

\begin{figure}

\resizebox{\hsize}{!}{\includegraphics{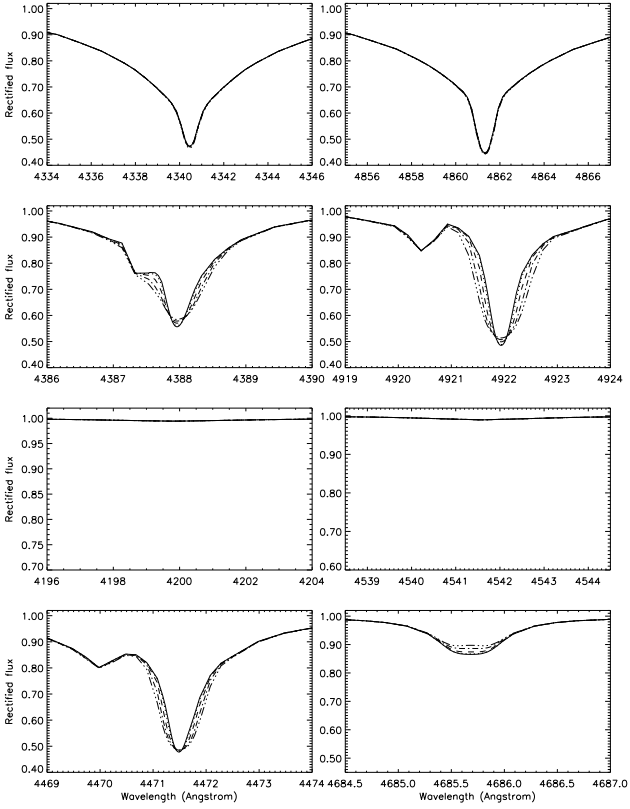}}
\caption[]{Same as Fig.~\ref{fig1}, but for $\log g$= 4.00, representing
late O and early B dwarfs.}
\label{fig2}

\end{figure}

\begin{table*}
\caption[ ]{Equivalent widths of the lines in Figs. \ref{fig1},
\ref{fig2} and \ref{fig3}, in m\AA. For every line the values for
microturbulence velocities of 0, 5, 10, 15 and 20 kms$^{-1}$ are given 
from top to bottom. Model
parameters in the first column are $T_\mathrm{eff}$ in K, $\log g$ in cgs units and $\epsilon$.}
\label{tab1}
\begin{tabular}{lcccccccc} 
\hline
Params. & H$_\gamma$ & H$_\beta$ & \ion{He}{i} 4387 & \ion{He}{i} 4922 & \ion{He}{i} 4471 & \ion{He}{ii} 4200 & \ion{He}{ii} 4541 & \ion{He}{ii} 4686 \\
\hline
30\,000 & 1895 & 2149 & 403 & 503 & 622 & 195 & 234 & 416 \\
3.05    & 1896 & 2150 & 415 & 526 & 637 & 195 & 234 & 418 \\
0.10    & 1896 & 2151 & 446 & 585 & 682 & 196 & 235 & 415 \\
	& 1904 & 2160 & 485 & 664 & 750 & 199 & 238 & 410 \\
	& 1904 & 2173 & 525 & 753 & 836 & 201 & 241 & 401 \\
\hline
30\,000 & 3831 & 3832 & 599 & 733 & 1158 & 42  & 50  & 161 \\
4.00    & 3832 & 3833 & 605 & 747 & 1167 & 42  & 50  & 160 \\
0.10    & 3834 & 3836 & 618 & 780 & 1189 & 42  & 50  & 157 \\
	& 3839 & 3843 & 636 & 825 & 1220 & 42  & 50  & 151 \\
	& 3850 & 3857 & 654 & 873 & 1252 & 42  & 50  & 145 \\
\hline
40\,000 & 2005 & 2393 & 120 & 238 & 391 & 491 & 614 & 685 \\
3.40    & 2004 & 2392 & 118 & 240 & 402 & 491 & 613 & 684 \\
0.10    & 2001 & 2385 & 118 & 255 & 434 & 492 & 613 & 685 \\
	& 2020 & 2406 & 120 & 279 & 480 & 519 & 646 & 667 \\
	& 2011 & 2392 & 120 & 299 & 530 & 522 & 649 & 669 \\
\hline
\end{tabular}
\end{table*}

\begin{figure}

\resizebox{\hsize}{!}{\includegraphics{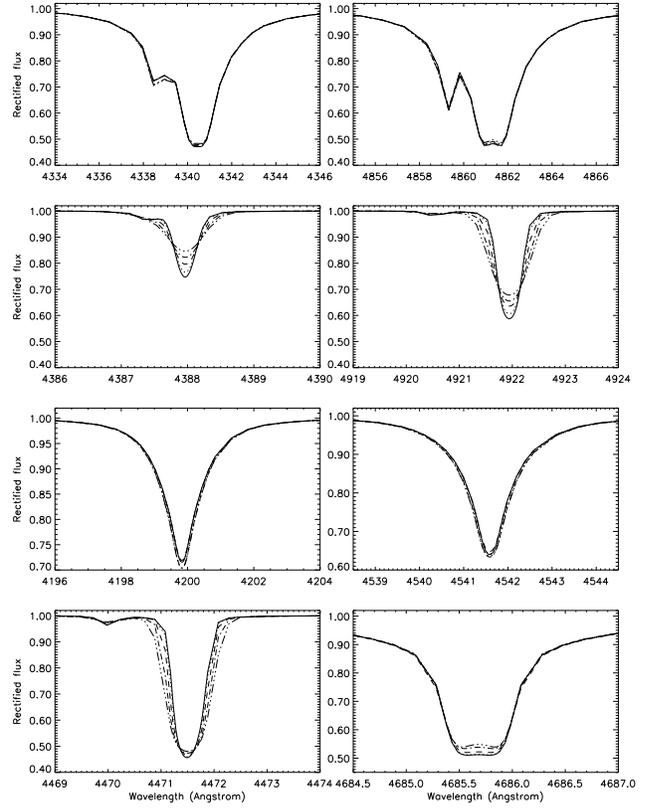}}
\caption[]{Same as Fig.~\ref{fig1}, but for $T_\mathrm{eff}$= 40\,000, $\log
g$= 3.40, $\epsilon$= 0.10, representing early O supergiants}
\label{fig3}

\end{figure}

Figs.~\ref{fig1}, \ref{fig2} and \ref{fig3} display the behaviour of H
and He line profiles with microturbulence, for three different models,
adequate respectively for late O supergiants, late O dwarfs 
and early O supergiants.  In Table ~\ref{tab1} equivalent widths are given for all these lines. The first two Figs. will allow us to compare the
effects of microturbulence when varying $\log g$, 
whereas the first and last Figs. will be used to compare the effects when
varying $T_\mathrm{eff}$ . The other calculated models
behave consistently with what it is shown in Figs.~\ref{fig1}--
\ref{fig3} and will not be further discussed.

Looking first at Fig.~\ref{fig1} we see that there are two different line
groups. The first one is composed by the strong H lines and the relatively
weak lines \ion{He}{ii} $\lambda\lambda$ 4200, 4541. All these lines are not
affected by microturbulence, although our highest values are comparable to
the thermal velocity of H atoms and well above that of He atoms. The reason
is that the Stark broadening dominates the profiles, and therefore masks the
changes that microturbulence produce.

The second group is composed by \ion{He}{i} lines and the strong
\ion{He}{ii}\,$\lambda$4686 line. \ion{He}{i} lines show sensitivity to microturbulence in
both core and wings, while the core of \ion{He}{ii} $\lambda$4686 is desaturated by
microturbulence, but its effect is masked in the wings by the Stark
broadening, because, as it has been explained by Smith \& Howarth
(\cite{Smith&Howarth98}), the effect of microturbulence will depend on the
stepness of the line wings.

When we then compare Figs.~\ref{fig1} and \ref{fig2} we see that
 the increased pressure broadening (directly related to
the increased density) simply reduces the effect of microturbulence
in the line profiles. Thus, as gravity increases, the effect of
microturbulence decreases for all considered lines. In the case of
 \ion{He}{ii} lines this effect is reinforced by the displacement
introduced in the He ionization equilibrium, that makes \ion{He}{ii}
lines weaker.

More interesting is the comparison with higher temperatures.  Looking to
Fig.~\ref{fig1} and Fig.~\ref{fig3} we can see the behaviour of the lines of
supergiants with increasing $T_\mathrm{eff}$. For higher values of
$T_\mathrm{eff}$ marginal effects on \ion{H}{i} and \ion{He}{ii} line cores
are seen (Fig.~\ref{fig3}), but only \ion{He}{i} lines and \ion{He}{ii}
$\lambda$4686 are again really sensitive to microturbulence. 
\ion{He}{ii} line profiles are still insensitive,
although they are now much stronger and even have a larger equivalent width
than \ion{He}{i} lines, because they are dominated by the Stark
broadening. The reason why \ion{He}{i} line profiles are still sensitive to microturbulence, although they
are much weaker, has to be found in the influence of microturbulence
on the shape of the absorption profile of weak lines. Thus, we see that
\ion{He}{i} $\lambda\lambda$ 4387, 4922 are not saturated in
the whole temperature range. Their equivalent widths increase with increasing
microturbulence, until, as $T_\mathrm{eff}$ increases, they become weak
enough to have equivalent widths, but not line profiles, independent of
microturbulence. This happens at 40\,000 K for \ion{He}{i} $\lambda$ 4387
(see Table \ref{tab1}) and at 45\,000 K for \ion{He}{i} $\lambda$ 4922.  For
them, Doppler broadening is important enough to let microturbulence shape the
profiles appreciably in the whole temperature range. \ion{He}{i} $\lambda$
4471 is saturated at $T_\mathrm{eff}$ 30\,000 K, 
but it desaturates at $T_\mathrm{eff}$= 35\,000, and starts behaving like the
other \ion{He}{i} lines.

Finally, we point out that all these results also apply for models
with $\epsilon=0.20$, and that they are a natural extension to O stars of the
results of Mc Erlean
et al. (\cite{McErlean98}) and Smith \& Howarth (\cite{Smith&Howarth98}).  

\section{The influence of microturbulence in the determination of parameters
of O-stars}

In this work we are interested in estimating the changes introduced in the
derived stellar
parameters by the inclusion of microturbulence in the analysis of
O stars, with special attention to its influence in the derived He
abundance. As we saw in the previous section, models of low gravity, 
corresponding to supergiants, are more sensitive to microturbulence, and so we
expect supergiants to show the largest changes. 

Values of microturbulence of 10, 12 kms$^{-1}$ are found for late O and early
B stars (see references in the introduction), therefore we decided to choose
a fixed value for microturbulence of 15 kms$^{-1}$ to perform a test of the
influence of microturbulence in the determination of $T_\mathrm{eff}$ , $\log g$ and He
abundance.

As a first approach to the problem, we take a model spectrum at
$\xi$=15 kms$^{-1}$ as the \emph {observed} spectrum, to which we fit model 
profiles in a grid around it at $\xi$=0 kms$^{-1}$, with parameters differing
 in 500 to
1000 K in $T_\mathrm{eff}$, 0.05 to 0.10 in $\log g$ and 0.02 to 0.04 in
$\epsilon$ (these small values are suggested by previous inspection of a 
larger, coarser grid). To maximize the effect of microturbulence, we
consider neither rotational broadening nor instrumental broadening. 
Taking advantage of the fact that we are
using only theoretical profiles (although the one with $\xi$= 15 kms$^{-1}$
has been adopted as \emph {the observation}) we use a least--squares fit
procedure to determine the best fit.

For each line of each model in the grid, we calculate the
quadratic difference with the \emph {observed} line, and then we add the
results for all lines of a given ion. For example for H we calculate: 
$$
\chi_{\mathrm i}^2 (\mathrm H)= \sum_{H\ lines}^{} {\frac{1}{\Delta \lambda}}
\sum_{\lambda}^{} {w_{\lambda} (f_{\lambda,grid} -f_{\lambda,obs})}^2
$$
as the result for H for model i of the grid, where $w_{\lambda}$ is the
spectral sampling and $\Delta \lambda$ is the wavelength interval
containing the whole line.  We perform these calculations for all the models
in the grid and then normalize these values to their minimum, so that a 1.00
gives us the model that best fits the lines of the ion at all temperatures,
gravities and He abundances.  The calculation is performed in the same way
for He lines, taking \ion{He}{i} and \ion{He}{ii} lines separately.  Finally we define
for model i:

$\chi_{\mathrm i}^2$= $\chi_{\mathrm i}^2(\mathrm H)$+$\chi_{\mathrm i}^2$(\ion{He}{i})+$\chi_{\mathrm i}^2$(\ion{He}{ii})

so that the best fit is adopted to be
that of the model with the minimum value of $\chi_{\mathrm i}^2$. 

\begin{figure}
\resizebox{\hsize}{!}{\includegraphics{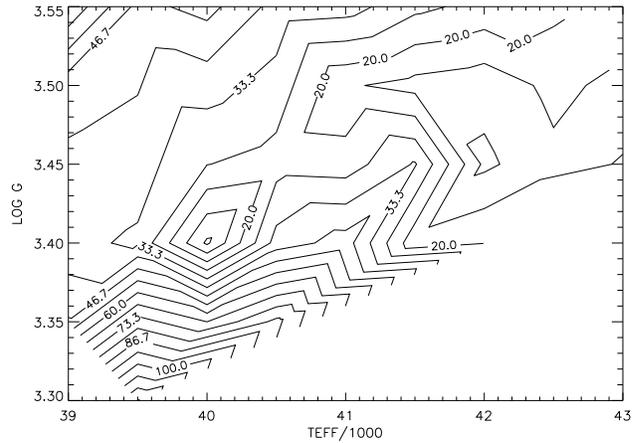}}
\caption[]{$\chi^2$-Contour levels in the $T_\mathrm{eff}$ -- $\log g$ plane for $\epsilon$= 0.10,
showing that the best fit for the {\emph {observed}} model 
at this abundance is still a model with the same parameters}
\label{contour}
\end{figure}

We have carried on the exercise for a model with $T_\mathrm{eff}$=40\,000 K,
$\log g$=3.40, $\epsilon$=0.10 and $\xi$=15 kms$^{-1}$ as the \emph {observed}
spectrum.  In Fig.~\ref{contour} we have plotted contour levels of $\chi^2$
in the $T_\mathrm{eff}$ -- $\log g$ plane for $\epsilon$ = 0.10.  We see
that, at this He abundance, there are two models that fit well the \emph
{observation}, with parameters $T_\mathrm{eff}$= 40\,000K, $\log g$= 3.40 and
$T_\mathrm{eff}$= 42\,000K, $\log g$= 3.45. Although the last model has a
slightly lower value of $\chi_{\mathrm i}^2$ (5.190), the dispersion in
the individual $\chi_{\mathrm i}^2$ values is larger, reflecting the fact
that looking at the line
fits we would choose the other as the best compromise. This is a indication of the
differences between both criteria, we should improve our $\chi^2$ fitting
criterium in order to match the fitting by eye.

\begin{table}
\label{squares}
\caption[ ]{$\chi_{\mathrm i}^2$ of models of the grid that best fit the {\emph {observed}} spectrum, with parameters $T_\mathrm{eff}$=40\,000 K, $\log g$=3.40, $\epsilon$=0.10 and $\xi$=15 km$s^{-1}$, see text for details.}
\label{tab2}
\begin{tabular}{ccccccc} 
\hline
{$T_\mathrm{eff}$} & {$\log g$} & {$\epsilon$} & $\chi_{\mathrm i}^2(\mathrm H)$ & $\chi_{\mathrm i}^2(\mathrm HeI)$ & $\chi_{\mathrm i}^2(\mathrm HeII)$ & $\chi_{\mathrm i}^2$ \\
\hline
41\,000 & 3.50 & 0.06 & 9.46 & 2.50 & 9.12 & 21.07 \\
41\,000 & 3.45 & 0.08 & 1.51 & 2.25 & 1.09 &  4.85 \\
40\,000 & 3.40 & 0.10 & 1.36 & 1.40 & 2.67 &  5.44 \\
42\,000 & 3.45 & 0.10 & 1.00 & 3.19 & 1.00 &  5.19 \\
39\,000 & 3.35 & 0.12 & 2.07 & 1.82 & 2.06 &  5.96 \\
40\,000 & 3.35 & 0.14 & 6.81 & 1.60 & 7.13 & 15.54 \\
\hline
\end{tabular}
\end{table}

\begin{figure}
\resizebox{\hsize}{!}{\includegraphics{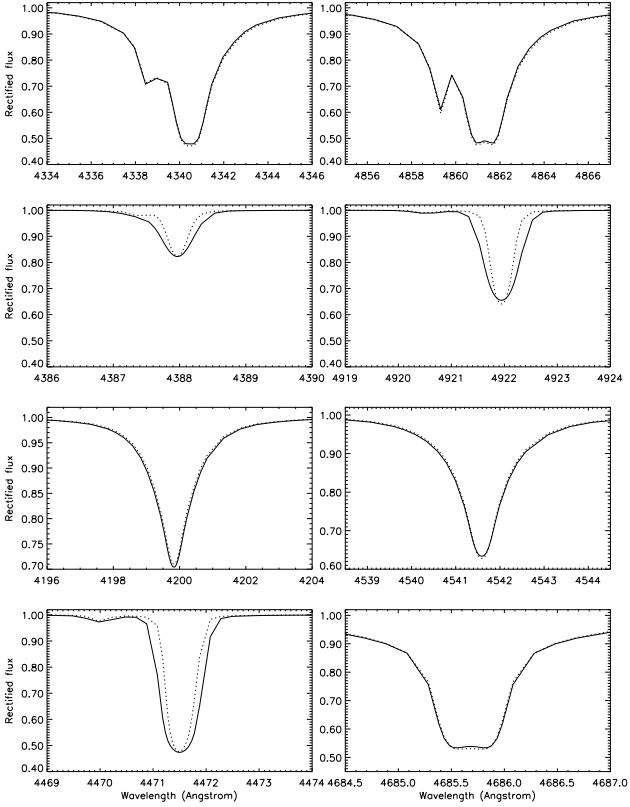}}
\caption[]{The fit of the ``best model'' (dotted, $T_\mathrm{eff}$= 41\,000 $\log g$= 3.45 
$\epsilon$= 0.08, $\xi$= 0 kms$^{-1}$ ) to the {\emph {observed}} one (solid, $T_\mathrm{eff}$= 40\,000 $\log g$= 3.40 $\epsilon$= 0.10, $\xi$= 15 kms$^{-1}$ ).}
\label{teor}
\end{figure}

In Table \ref{tab2} we list the results for the best fits at other He
abundances. We see inmediately that for $\epsilon$ = 0.06 and 0.14 we find
large values of $\chi^2$, indicating a relatively poor
fit. On the contrary, the models for $\epsilon$= 0.12, 0.10, 0.08 fit
with approximately the same quality (without a more detailed study
we cannot say whether the difference in $\chi^2$ is significant).
In a fit by eye we would choose the model at $\epsilon$ 0.08 as the
best fit. Actually, this is the model with the lowest value of $\chi^2$ 
and has a \emph {lower}
He abundance than the \emph {observed} spectrum. The reason is that
the larger gravity rather than the He abundance mimics the 
microturbulence effect. In principle, it would be also possible to choose
the model with $\epsilon$= 0.12, in which the higher He abundance
corresponds with a lower gravity. Fig.~\ref{teor} shows the fit of the
``best model'' to the \emph {observed} spectrum. The lack of rotation and
instrumental broadening makes the differences in \ion{He}{i} lines appear evident.

Thus, the conclusion of our exercise is that neglection of 
microturbulence will change slightly the derived stellar parameters, 
but keeping them within our standard error box (see next Sect.).
The direction in which parameters are moved will depend 
on the criteria used for defining a model as ``the best model fit'',
such as the relative weight given to different lines, or to the
core and wings of a line. Thus the definition of these
criteria for the ``best model'' will be a critical point
of any future automatic fitting procedure, which will soon
be demanded by new observing capabilities.

\section{Spectral analysis including microturbulence}

Now we want to see the difference we obtain in the derived parameters
of real observed O stars spectra when we include microturbulence in the
calculations. We selected two late-O supergiants to see how our results match
with previous works, as well as one intermediate and 
two early ones, in order to cover the whole O spectral type.
We chose \object{HD 5\,689} in particular, an O6 V fast rotator
with a high He-overabundance derived in previous works without
microturbulence (see Herrero et al., \cite{Herrero00}), to check how it can 
modify this strong overabundance.

The description of the observations and data reduction can be found in
Herrero et al. (\cite{Herrero92}, \cite{Herrero00}).

To analyse stellar spectra we first determine the projected rotational
velocity of the star (see Herrero et al. \cite{Herrero92} for details). 
This is an additional parameter that, in fact, reduces the effect
of microturbulence in the derivation of stellar parameters. The results of the
analyses are listed in Table \ref{tab3}.

The fitting procedure and adopted errors of $\pm$ 1\,500 K in
$T_\mathrm{eff}$, $\pm$ 0.1 in $\log g$ and $\pm$ 0.03 in $\epsilon$ are explained
in Herrero et al. (\cite{Herrero99}).

To make the new analyses with microturbulence 
we construct a model with the same parameters but for
$\xi$= 15 kms$^{-1}$, and also a small grid around it with changes in
$T_\mathrm{eff}$ of $\pm$ 500 to 2000 K, in $\log g$ of $\pm$ 0.05 to 0.15 dex
and
in $\epsilon$ of $\pm$ 0.02 to 0.04. These small changes are suggested by the
results of the preceeding section, and are in fact confirmed by the
small differences in the fit we see between the two models at 0 and 15
kms$^{-1}$. The new best fit is found as explained above, and it gives the
new parameters for the star at $\xi$=15 kms$^{-1}$. 

All the stars except \object{HDE\,338\,926} have been previously analyzed by 
our group without
microturbulence (Herrero et al., \cite{Herrero92}, \cite{Herrero00}).
In the present analysis, small differences for the
stars in common, specially with
respect to the first reference, can be found due to the larger weight given
here to \ion{He}{i} $\lambda$4387, rather than to \ion{He}{i} $\lambda$4922,  and to the inclusion of
line--blocking in the calculations (the temperatures of the late
type supergiants are slightly hotter in the present analysis).
All this also helps to solve the small difference between \ion{He}{i} $\lambda$ 4387 and
\ion{He}{i} $\lambda$ 4922 found in Herrero et al. (\cite{Herrero92}) for these
late O supergiants.

\begin{figure}
\resizebox{\hsize}{!}{\includegraphics{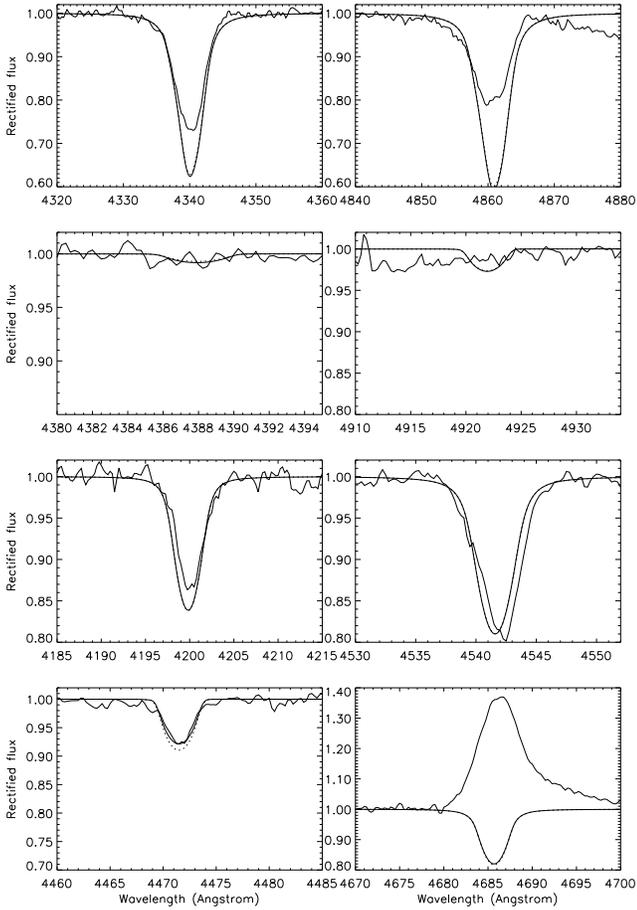}}
\caption[]{Fits with $\xi=0$ kms$^{-1}$ (solid line) and $\xi=15$ kms$^{-1}$
(dotted) to the star \object{HD 14\,947}. The parameters are $T_\mathrm{eff}$=
45\,000, $\log g$= 3.50, $\epsilon$= 0.15 at $\xi=0$ kms$^{-1}$ and the same
at 15 km$^{-1}$, which are the converged models closest to the final adopted parameters, with $\log g$ = 3.45}
\label{fig4a}
\end{figure}

\begin{figure}
\resizebox{\hsize}{!}{\includegraphics{H1917.f7}}
\caption[]{Fits with $\xi=0$ kms$^{-1}$ (solid line) and $\xi=15$ kms$^{-1}$
(dotted) to the star \object{HD 5\,689}. The parameters are $T_\mathrm{eff}$=
40\,000, $\log g$= 3.40, $\epsilon$= 0.25 at $\xi=0$ kms$^{-1}$ and
$T_\mathrm{eff}$= 40\,000, $\log g$= 3.35, $\epsilon$= 0.22 at $\xi=15$
kms$^{-1}$}
\label{fig4}
\end{figure}

\begin{figure}
\resizebox{\hsize}{!}{\includegraphics{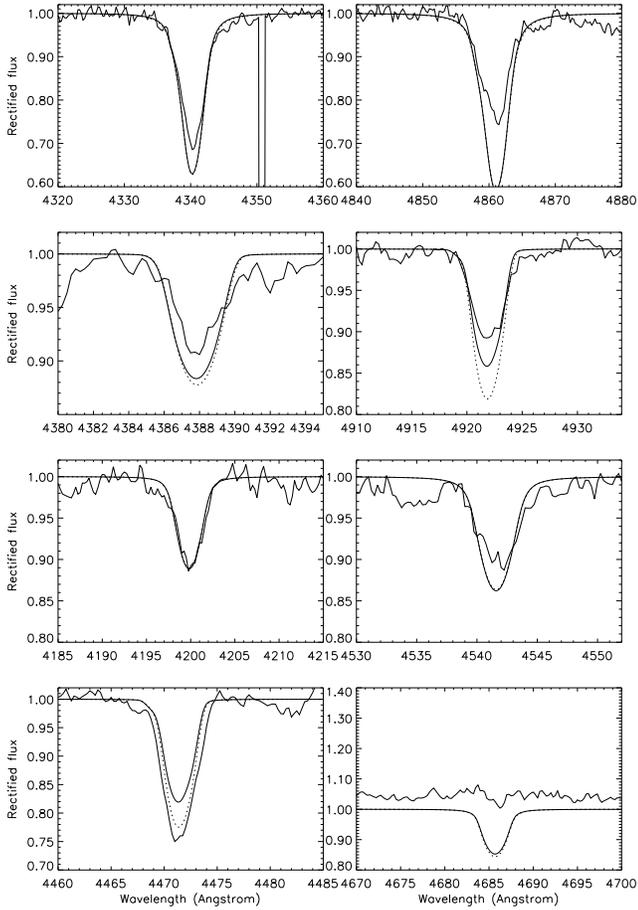}}
\caption[]{Fits with $\xi=0$ kms$^{-1}$ (solid line) and $\xi=15$ kms$^{-1}$
(dotted) to the star \object{HDE 338\,926}. The parameters are $T_\mathrm{eff}$=34\,000, $\log g$= 3.05, $\epsilon$= 0.15 at $\xi=0$ kms$^{-1}$ and
$T_\mathrm{eff}$= 34\,500, $\log g$= 3.10, $\epsilon$= 0.13 at $\xi=15$
kms$^{-1}$. Again these are the closest models that could be converged, for a  final adopted $\log g$ of 3.00}
\label{fig4b}
\end{figure}

\begin{figure}
\resizebox{\hsize}{!}{\includegraphics{H1917.f9}}
\caption[]{Fits with $\xi=0$ kms$^{-1}$ (solid line) and $\xi=15$ kms$^{-1}$ (dotted) to the star \object{HD 210\,809}. The parameters are $T_\mathrm{eff}$=33\,500, $\log g$=3.10, $\epsilon$=0.10 at $\xi=0$ kms$^{-1}$ and $T_\mathrm{eff}$=34\,500, $\log g$=3.15, $\epsilon$=0.08 at $\xi=15$ kms$^{-1}$}
\label{fig5}
\end{figure}

\begin{figure}
\resizebox{\hsize}{!}{\includegraphics{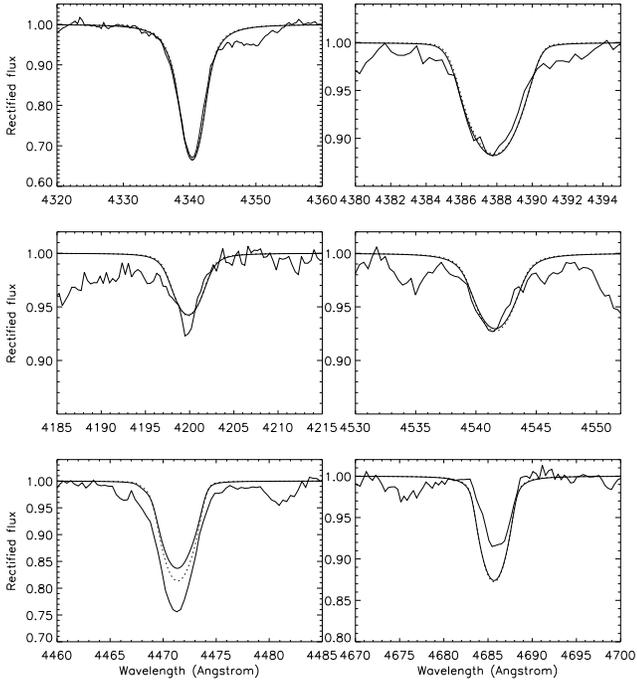}}
\vspace{-3cm}
\caption[]{Fits with $\xi=0$ kms$^{-1}$ (solid line) and $\xi=15$ kms$^{-1}$
(dotted) to the star \object{HD18\,409}. The parameters are $T_\mathrm{eff}$=
31\,500, $\log g$= 3.10, $\epsilon$= 0.11 at $\xi=0$ kms$^{-1}$ and
$T_\mathrm{eff}$= 32\,000, $\log g$= 3.10, $\epsilon$= 0.09 at $\xi=15$
kms$^{-1}$. We could not use neither H$_\beta$ nor \ion{He}{i} $\lambda$ 4922
for the fit, as we don't have observations of this star in this range.}
\label{fig6}
\end{figure}

In Table \ref{tab3} we see the results for the five stars. Changes in the
parameters induced by microturbulence are not beyond the standard error box
of the analysis 
, and in particular, we see that He abundance
is reduced for four of the stars, but only slightly. 
In Table \ref{tab4} we see the values obtained for the radius, mass
and luminosity, following the same procedures outlined in
Herrero et al. (\cite{Herrero92}). We see that changes are not significant.
(We point out that the values for mass, radius and luminosity given
in Table \ref{tab4} for \object{HD\,210\,809} and \object{HD\,18\,409} differ slightly from those
given in Herrero et al. (\cite{Herrero92}). These differences are not due
to the new parameters, but to a change in the distance moduli.
Herrero et al. used the values from Humphreys (\cite{hum78}), while
we use the values from Garmany \& Stencel (\cite{GarS})).

\begin{table*}
\caption[ ]{Stellar parameters with values of microturbulence of 0 and 15
kms$^{-1}$. Effective temperatures are given in K and surface gravities in
c.g.s. units. Spectral classifications 
are taken from Walborn (\cite{wal73}), except that of
\object{HD\,5\,689}, which is taken from Garmany \& Vacca (\cite{gv91}),
and that of \object{HDE\,338\,926}, which is taken from Humphreys (\cite{hum78}; 
Humpreys lists this star as BD+24 3866). 
Some final parameters are extrapolated
as we could not converge the model with the indicated values of $\log g$
but others with $\log g$ 0.05 larger, with all other parameters the same. 
This is indicated by giving the parameters in italics }
\label{tab3}
\begin{tabular}{llccccccc} 
\hline
Star & Clasif. & $V_\mathrm{r} \sin i$ & \multicolumn{2}{c}{$T_\mathrm{eff}$} & \multicolumn{2}{c}{$\log g$} & \multicolumn{2}{c}{$\epsilon$} \\
     &    & kms $^{-1}$ & 0 kms$^{-1}$ & 15 kms$^{-1}$ & 0 kms$^{-1}$ & 15 kms$^{-1}$ & 0 kms$^{-1}$ & 15 kms$^{-1}$ \\
\hline
\object{HD 14\,947}  & O5 If$^{+}$ & 140 & 45\,000 & {\it 45\,000} & 3.50 & {\it 3.45} & 0.15 & {\it 0.15} \\
\object{HD 5\,689}   & O6 V   & 250 & 40\,000 & 40\,000 & 3.40 & 3.35 & 0.25 & 0.22 \\
\object{HDE 338\,926} & O8 f   & 130 & {\it 34\,000} & {\it 34\,500} & {\it
3.00} & {\it 3.00} & {\it 0.15} & {\it 0.13} \\
\object{HD 210\,809} & O9 Iab & 120 & 33\,500 & 34\,500 & 3.10 & 3.15 & 0.10 & 0.08 \\ 
\object{HD 18\,409}  & O9.7 Ib& 160 & 31\,500 & 32\,000 & 3.10 & 3.10 & 0.11 & 0.09 \\ 
\hline
\end{tabular}
\end{table*}

\begin{table*}
\caption[ ]{Radius, mass and luminosity for the analyzed objects. As in 
Table \ref{tab3}, italic numbers mean values obtained by extrapolation }
\label{tab4}
\begin{tabular}{lcccccc} 
\hline
Star & \multicolumn{2}{c}{$R/R_{\sun}$} & \multicolumn{2}{c}{$M/M_{\sun}$} & \multicolumn{2}{c}{$\log (L/L_{\sun})$} \\
     & 0 kms$^{-1}$ & 15 kms$^{-1}$ & 0 kms$^{-1}$ & 15 kms$^{-1}$ & 0 kms$^{-1}$ & 15 kms$^{-1}$ \\
\hline
\object{HD 14\,947}  & 14.8 & {\it 14.9} & 26.5 & {\it 24.1} & 5.91 & {\it 5.91} \\
\object{HD 5\,689}   &  7.7 &  7.8 &  7.8 &  7.4 & 5.13 & 5.14 \\
\object{HDE 338\,926} & {\it 24.4} & {\it 24.3} & {\it 23.8} & {\it 23.5} &
{\it 5.86} & {\it 5.88} \\
\object{HD 210\,809} & 18.2 & 17.9 & 16.0 & 17.4 & 5.57 & 5.61 \\ 
\object{HD 18\,409}  & 18.8 & 18.8 & 18.7 & 18.6 & 5.50 & 5.52 \\ 
\hline
\end{tabular}
\end{table*}

In Figs. ~\ref{fig4a} to ~\ref{fig6} we can see the fits with microturbulence
0 and 15 kms$^{-1}$. The quality of the fits to individual lines is the same
at both values, although the triplet line 
\ion{He}{i} $\lambda$ 4471 becomes stronger at 15 kms$^{-1}$, which
produces an improved fit, but not to the
extent as to completely resolve the so--called dilution effect. Sometimes,
a compromise between \ion{He}{i} $\lambda$4387 and \ion{He}{i} $\lambda$4922
is adopted, like in \object{HD\,5\,689} and \object{HD\,210\,809} (realize 
that for 15 km$s^{-1}$ the first 
one is slightly weak and the second one slightly strong).

With respect to \object{HDE\,338\,926}, that has been analyzed here for the first
time, we have to point out that the analysis has been very difficult.
The final parameters are extrapolated beyond the models we
could converge. However, we are confident that these parameters do
characterize the star appropriately (or as appropriately as those of
other stars), because the fits with already
converged models are reasonably good, and because we are able to fit the star
with converged models when we do not consider line--blocking. Thus,
it is only the small temperature increase introduced by line--blocking 
that moves the star beyond the convergence region. 
The value we obtain for the 
evolutionary mass, derived from the tracks by Schaller et al., 
(\cite{Schall92}), is 57.6 $M_{\sun}$ without microturbulence, and
56.8 $M_{\sun}$ with 15 kms$^{-1}$. Comparing with the values in
Table \ref{tab4}, it is clear that \object{HDE\,338\,926} also 
shows a mass discrepancy, as usual. Of course, we
will have to reanalyze \object{HDE\,338\,926} in the future with spherical,
mass lossing models.

\section{Discussion}

We will not go here into the discussion of the real physical entity of
microturbulence, neither on the validity of the aproximation of small scale
turbulent movements assumed to introduce it just as an extra Doppler width,
which may be not suitable for the big values we deal with. This is beyond the
scope of this work. As explained in Sect. 2 we do not introduce
microturbulence in the structure calculations because we think that,
having neglected other motions, it is actually not more physically
consistent to introduce a turbulent pressure term.
We just accept the necessity of using microturbulence
in the analysis of stellar spectra,
specially in the determination of metal abundances, that is our final
interest. 

We find that contrary to our previous considerations, \ion{He}{i} lines and
\ion{He}{ii} $\lambda$ 4686 do have profiles sensitive to the usual
values of microturbulence found in OB stars. We confirm this for the
whole range of parameters describing O and early B spectra, in agreement with
McErlean et al. (\cite{McErlean98}) and Smith \& Howarth (\cite{Smith&Howarth98})
for early B and late O supergiants respectively. This sensitivity is first
due to the fact that the Stark broadening is not dominating completely these
profiles and second to the high values of microturbulence involved, that are
comparable to the thermal velocity of He ions. Taking (0.84 $T_\mathrm{eff}$) as 
representative of the temperature in the zone
of formation of the lines, we find that for $T_\mathrm{eff}$= 30\,000 K
$v_\mathrm{th}$ of He ions is 10 kms$^{-1}$, and for $T_\mathrm{eff}$= 45\,000
K it is 12 kms$^{-1}$.
For H ions, with a fourth of the He atomic weigth, 
thermal velocities are two times larger,
so thermal broadening and Stark broadening dominate the
profile. For all other \ion{He}{ii} lines Stark broadening is hiding the
effect of microturbulence.

Quantifying the effect of microturbulence on the determination of stellar
parameters we find that they are not changed beyond the standard error box of
our analyses. This implies that for stars with high He overabundances, such
as \object{HD 5\,689} analysed here, the {\it He discrepancy} will not be solved by
considering microturbulence. This result is also supported by what we
obtained in Sect. 3.

This seems to be in contradiction with Smith \& Howarth
(\cite{Smith&Howarth98}) and McErlean et al. (\cite{McErlean98}), who affirm
that solar He abundances are found for the supergiants they analyse,
when microturbulence is considered. However, the situation is still unclear,
as an analysis of published results can show.

Let us begin with the He overabundances obtained by Herrero et al.
(\cite{Herrero92}).  Following the argument by Smith \& Howarth
(\cite{Smith&Howarth98}) and McErlean et al. (\cite{McErlean98}), the
preferred use of the strong line \ion{He}{i} $\lambda$ 4922 in the analyses
(slightly more sensitive to microturbulence than \ion{He}{i} $\lambda$
4387) would be responsible for the high He abundances found in that work. As
a conclusion, the derived He overabundances would be an artifact introduced
by the neglection of microturbulence.

This argument applies in the cases in which Herrero et al. 
(\cite{Herrero92}) found a discrepancy between \ion{He}{i}
$\lambda$ 4922 and \ion{He}{i} $\lambda$ 4387, but 
not in the rest of the cases.

\begin{figure}

\resizebox{\hsize}{!}{\includegraphics{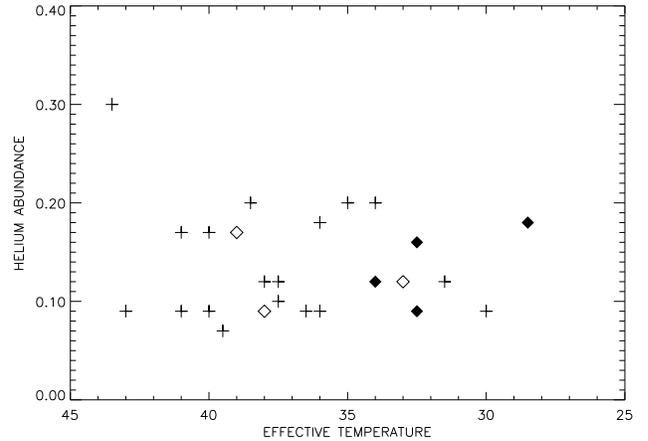}}
\caption[]{$\epsilon$ versus $T_\mathrm{eff}$ (in thousands of Kelvin) in the sample
of Herrero et al. (\cite{Herrero92}). Filled diamonds represent stars for
which Herrero et al. report a large discrepancy between  
\ion{He}{i} $\lambda$ 4922 and \ion{He}{i} $\lambda$ 4387. Open diamonds
represent stars for which Herrero et al. report a moderate discrepancy
between these two lines. Plus signs represent stars for which He\-rre\-ro et al.
do not report a discrepancy}
\label{fig10}

\end{figure}

In Fig.~\ref{fig10} we have plotted He abundance versus $T_\mathrm{eff}$, as
derived by Herrero et al. (\cite{Herrero92}), using different symbols for
stars for which a discrepancy between these two
lines was reported. We can see that the priority given to \ion{He}{i} $\lambda$ 4922 can
affect some of the abundances derived, specially at lower temperatures.

To illustrate the situation, let us briefly 
discuss one of these objects for which
Herrero et al. report a difference between \ion{He}{i} $\lambda$ 4922 and
\ion{He}{i} $\lambda$ 4387: \object{HD\,210\,809}. This star 
has been analyzed here again,
where we have given more weight to \ion{He}{i} $\lambda$ 4387 (contrary to
Herrero et al.). The final difference between the He abundance obtained here
with a microturbulence of 15 kms$^{-1}$ and that obtained by Herrero et
al. is 0.04 (our new He abundance being lower). This, however, has to be
adscribed to different effects. First, giving more weight to \ion{He}{i}
$\lambda$ 4387 in absence of microturbulence results in a higher 
temperature (as already stated by Herrero
et al.), which leads to a lower He abundance to keep the fit of \ion{He}{ii}
lines. The effect of line blocking, not included in Herrero et al., goes in
the same direction. This accounts for a difference of 0.02 and the additional
difference of 0.02 is purely due to microturbulence (see Table \ref{tab3}).
Comparing our Fig.~\ref{fig5} with Fig. 5 of Herrero et al. 
(\cite{Herrero92}) we see that 
the effects considered here help to 
reduce the discrepancy between the two \ion{He}{i} lines.

The discrepancy between \ion{He}{i} $\lambda$ 4387 and
\ion{He}{i} $\lambda$ 4922 found by Herrero et al. (\cite{Herrero92})
is larger at lower temperatures, and thus, around 30\,000 K it cannot
be solved by varying $T_\mathrm{eff}$ or introducing line blocking. Therefore,
around or below this temperature microturbulence is at present the only 
considered effect that could bring them into agreement.

In the literature we can find analyses of early B and late O
supegiants including microturbulence (Gies \& Lambert, \cite{Gies&Lambert92},
Smith \& Howarth, \cite{Smith&Howarth94},
\cite{Smith&Howarth98}, Smith et al., \cite{Smithetal98}, 
McErlean et al., \cite{McErlean98}). In Table \ref{tab5} we list
stars for which parameters have been derived with and without microturbulence.

\begin{table*}
\caption[ ]{OB supergiants whith parameters determined with and without 
microturbulence. Values of microturbulence vary between 10 and 15 kms$^{-1}$.
References are as follows: (1) Herrero et al. (\cite{Herrero00}); 
(2) this work; (3) Herrero et al. (\cite{Herrero92}); 
(4) Smith \& Howarth (\cite{Smith&Howarth94}); 
(5) Smith et al. (\cite{Smithetal98}); 
(6) Smith \& Howarth (\cite{Smith&Howarth98}); 
(7) Lennon et al. (\cite{Lennon91}); (8) McErlean et al. 
(\cite{McErlean98})}
\label{tab5}
\begin{tabular}{llccccccccc} 
\hline
Star & Clasif. & $V_\mathrm{r} \sin i$ & \multicolumn{2}{c}{$T_\mathrm{eff}$}
& \multicolumn{2}{c}{$\log g$} & \multicolumn{2}{c}{$\epsilon$} & 
\multicolumn{2}{c}{Ref.} \\
     &    & kms $^{-1}$ & $\xi=$ 0 & $\xi\ne$ 0 & $\xi=$ 0 & $\xi\ne$ 0 & $\xi=$ 0 & $\xi\ne$ 0 & $\xi=$ 0 & $\xi\ne$ 0 \\
\hline
\object{HD\,14\,947}  & O5 If$^+$& 140& 45.0 & 45.0 & 3.50 & 3.45 & 0.15 & 0.15 & 1 & 2 \\
\object{HD\,5\,689}   & O6 V     & 250& 40.0 & 40.0 & 3.40 & 3.35 & 0.25 & 0.22 & 1 & 2 \\
\object{HD\,210\,809} & O9 Iab   & 100& 33.0 & 34.5 & 3.10 & 3.15 & 0.12 & 0.08 & 3 & 2 \\
\object{HD\,154\,368} & O9.5 Iab & 85 & 33.0 & 32.0 & 3.07 & 3.00 & 0.13 & 0.13 & 4 & 5 \\
\object{HD\,123\,008} & ON9.5 Iab& 90 & 33.5 & 33.0 & 3.07 & 3.05 & 0.17 & 0.15 & 4 & 5 \\
\object{HD\,152\,003} & O9.7 Iab & 80 & 30.8 & 29.7 & 2.90 & 3.10 & 0.09 & 0.09 & 6 & 6 \\
	     &          &    &      & 30.5 &      & 3.00 &      & 0.09 &   & 5 \\ 
\object{HD\,18\,409}  & O9.7 Ib  & 160& 31.0 & 32.0 & 3.10 & 3.10 & 0.12 & 0.09 & 3 & 2 \\
\object{HD\,154\,811} & OC9.7 Iab& 125& 31.5 & 31.0 & 3.15 & 3.10 & 0.09 & 0.09 & 4 & 5 \\
\object{$\kappa$ Ori} & B0.5 Ia  & 80 & 25.0 & 27.5 & 2.70 & 3.00 & 0.20 & 0.10 & 7 & 8 \\
\hline
\end{tabular}
\end{table*}

Comparisons in Table \ref{tab5} have to be done with care. It mixes results
from different authors and different criteria. We see however
that it indicates that the changes found here
 are consistent with those of other authors, i.e.,
stellar parameters are changed within the uncertainties adopted here.
These changes do not follow a clear, systematic pattern (i.e., going
always in the same direction when introducing microturbulence) and thus
we conclude that the effect of microturbulence is indeed not larger
than present--day uncertainties. An exception to this might be
\object{$\kappa$ Ori}, which shows a large change in He abundance. However,
this large reduction of the He abundance found by McErlean et al.
(\cite{McErlean98}) as compared to Lennon et al.
(\cite{Lennon91}) is accompanied by a large change in the
stellar parameters. Even if we attribute the whole change to the
effect of microturbulence (and not to a new analysis with new data and
more refined calculations) we see that \object{$\kappa$ Ori} is by far the coolest
of the stars in Tables~\ref{tab3} and \ref{tab5} and thus we expect a larger 
influence of microturbulence for it. In addition, we note that McErlean et al.
(\cite{McErlean98}) do not exclude a larger He abundance for this star.
We conclude that data in the literature do not lead to the
conclusion that the {\it He discrepancy} in O and early B supergiants
is completely due to microturbulence, although microturbulence helps
to reduce it.

Except for the discussed reduction of the He abundance,
we do not see a clear pattern in the parameter changes in Tables~\ref{tab3}
and \ref{tab5}. $T_\mathrm{eff}$ and $\log g$ can both increase or decrease when 
microturbulence is introduced, and thus we are tempted to conclude that these
changes are still dominated by internal inconsistencies in the analyses
that appear when we compare values that cannot be distinguished within
the adopted uncertainties.

About the problem of the consistency of the fits to \ion{He}{i} lines, we
find that the dilution of \ion{He}{i} $\lambda$ 4471 is only partially
solved, even when line--blocking is considered as here. The fits to the rest
of \ion{He}{i} lines do not improve much either in the stars we
analyze here. So the consideration of
both microturbulence and line--blocking in the analysis cannot completely
make an agreement between the results from triplet and singlet \ion{He}{i}
lines, though they help to improve it.

\section{Conclusions}

We study for the first time the effect of microturbulence in the whole O
spectral range, from late to early O types. Introducing microturbulence in the
solution of the statistical equilibrium and transfer equations, and then in
the formal solution (i.e., in the absorption coefficient),
we conclude that for higher
gravities the effect in the lines is negligible. This together with the low
values of microturbulence usually found for high luminosity class stars, lead
us to conclude that only O supergiants have sensitivity to
microturbulence effects at a given $T_\mathrm{eff}$.

In examining the behaviour with microturbulence of the lines we use in our
analysis, we show that only \ion{He}{i} lines and the core of \ion{He}{ii}
$\lambda$ 4686 \AA~ are sensitive to microturbulence. For \ion{He}{i} lines we
show that, as should be expected, there is not a constant pattern for each
line, but it depends on the parameters considered, that determine the
strength of the line and its degree of saturation. This invalidates
generalizations to the whole O spectral range made upon results obtained just
for a certain spectral type.

In order to quantify the sensitivity of stellar parameters to microturbulence
we find that changes in the parameters induced by a value of microturbulence
of 15 kms$^{-1}$ are enclosed within the standard error box of our
analyses. We think that the lack of a clear pattern in the changes induced
in $T_\mathrm{eff}$ and $\log g$ is just due to the fact that we are varying the
parameters within this error box. We do however find a systematic change
in $\epsilon$ towards lower He abundances when microturbulence is introduced.
In particular we find that late O supergiants show a decrease of
0.02--0.04 in $\epsilon$ (this last value when including other effects that 
add to microturbulence), which is in agreement with previous results
pointing to the inverse relation between the microturbulence assumed and the
He abundance obtained (Smith \& Howarth \cite{Smith&Howarth98}).
Early types are less sensitive to microturbulence, and might not show
a difference in the derived He abundance.

Thus microturbulence is not capable of explaining the {\it He discrepancy} at
all for early O stars, and neither it is for late O types with high
overabundances.

Looking to individual lines we find that the fits to \ion{He}{i} $\lambda$
4471 \AA~ are improved when considering microturbulence, but not to the
extent of completely explaining its dilution. On the other hand \ion{He}{i}
$\lambda$$\lambda$ 4922, 4387 \AA~ are sometimes slightly better and sometimes
slightly worse fitted, in the last case with model cores
a bit too strong or a bit too weak, respectively. The
rest of the lines keep the same quality in the fit. The {\emph{\ion{He}{i} lines
problem}} is therefore only partially solved by simply considering
microturbulence, even with line--blocking included in the model profiles.

Therefore our conclusion is that microturbulence is affecting the derivation
of stellar parameters, but its effect
is comparable to the adopted uncertainties. Thus it can reduce
moderate He overabundances and solve line fit quality differences, but it
cannot explain by itself large He overabundances in O stars, and we are forced
to conclude that these are due to other effects, whether real or
caused by artifacts in our analyses. This last point will probably not find a 
definitive answer until we are able to derive reliable abundances of
C, N and O in the atmospheres of O stars that we can correlate with
the He abundances.

\begin{acknowledgements}
We want to thank Neil McErlean and Danny Lennon for their help, suggestions  and
many clarifying discussions. AH wants to acknowledge
support for this work by the spanish DGES under project PB97-1438-C02-01
\end{acknowledgements}



\end{document}